\documentclass[10pt]{article}
\usepackage{amsmath}
\usepackage{amssymb}

\usepackage{graphicx}

\usepackage{color} 

\usepackage[labelfont=bf,labelsep=period,justification=raggedright]{caption}

\usepackage[sorting=none]{biblatex}
\bibliography{ia}

\makeatletter
\renewcommand{\@biblabel}[1]{\quad#1.}
\makeatother

\date{}

\usepackage{url}

\begin{document}

\begin{flushleft}
{
\Large
\textbf{How to Create an Innovation Accelerator}\\
}
Dirk Helbing$^{1,2}$ and Stefano Balietti$^{1}$ 
\\[3mm]
\bf{1} ETH Zurich, CLU, Clausiusstr. 50, 8092 Zurich, Switzerland
\\
\bf{2} Santa Fe Institute, 1399 Hyde Park Road, Santa Fe, NM 87501, USA
\\
$\ast$ E-mail: 
{\normalfont dhelbing@ethz.ch and sbalietti@ethz.ch}
\end{flushleft}

\section*{Abstract}

The purpose of this White Paper of the EU Support Action ``Visioneer''
(see \url{www.visioneer.ethz.ch}) is to address the following goals:

\begin{enumerate}
\item Identify new ways of publishing, evaluating, and reporting
  scientific progress.
\item Promote ICT solutions to increase the awareness of new emerging
  trends.
\item Invent tools to enhance Europe's innovation potential.
\item Develop new strategies to support a sustainable technological
  development.
\item Lay the foundations for new ways to reach societal benefits and
  respond to industrial needs using ICT.
\end{enumerate}

\section{Introduction}

The way in which science is organized today has largely missed to use
the opportunities of the information revolution, particularly the Web2.0. It
will therefore be discussed which parts of scientific knowledge
creation and spreading need to be improved or reinvented, what tools are
available, and which ones need to be created to achieve the required
changes.
\par
Problems have become apparent in particular in the social and economic sciences,
which are facing emerging challenges at an accelerating rate. 
President Lee C. Bollinger of New York's prestigious Columbia University
described the situation as follows: ``The forces affecting societies around the world ... are powerful and
novel. The spread of global market systems ... are ... reshaping our world ..., raising profound questions.
These questions call for the kinds of analyses and understandings that academic institutions are uniquely capable of providing. Too many policy failures are fundamentally failures of knowledge'' \cite{Bollinger}. This has become particularly apparent during the recent financial and economic crisis, which is questioning
the validity of mainstream scholarly paradigms. Given the impact that the financial crisis has had on economies 
and societies all over the world---and will have for many more years---it appears necessary to get from a situation
of doing the aftermath of crises into a position of being able to anticipate and mitigate them efficiently, which also  
calls for contingency plans and the exploration of alternatives \cite{VisioneeSS}. 
\par
Altogether, this requires to close a number of
knowledge gaps and to accelerate the rate of knowledge creation, taking into account the wisdom obtained in other research fields. In other words, it seems appropriate to pursue a multi-disciplinary approach, involving researchers and methods from other disciplines, and to establish new institutional settings which remove or reduce obstacles impeding efficient knowledge creation. While the first part of this White Paper will make suggestions on how to modernize and improve the academic publication system, the second part will address the issue of supporting scientific coordination, communication, and co-creation in large-scale multi-disciplinary projects. Both constitute important elements of what we envision to be an ``Innovation Accelerator'' or ``Knowledge Accelerator''. 

\section{Identify new ways of publishing, evaluating and reporting
  scientific progress}\label{QJournal}
  
\subsection{Stylized characterization of the current situation}  
  
The way in which the current publication system works dates back to times where
\begin{enumerate}
\item scientists were a small and hand-selected elite,
\item the creation of papers was a cumbersome and slow process (using type-writers),
\item the publication of papers was quite expensive,
\item the world was changing at a relatively slow pace as compared to today,
\item coherent scientific paradigms could only be reached by a selection process that may be compared with cultivating a garden or forest (planting certain flowers or trees and removing others that did not fit well).
\end{enumerate}
Before we describe the paradigm shift that current and future information system will bring about regarding the way in which we create, disseminate, select, and harvest knowledge, let us start with highlighting some side effects that the present publication system seems to have (of course, not always, but tendentially):
The conventional way of spreading scientific
knowledge is passing through the system of peer-reviewed journals\footnote{At least this is valid for the natural sciences and part of the social sciences, in the humanities still book production is important \cite{Jasist}}. This means that the publication process is slow and that valuable information may be lost in the selection process. Surprisingly, despite a considerable literature, there is little sound
peer-review research examining criteria or strategies
for improving this process \cite{PeerReviewMarsh}. 

Before introducing the concepts of a Science 2.0 framework (Sec. \ref{Science2}) and of the Innovation Accelerator (Sec. \ref{howto_IA}), let us first discuss some problems of the selection process, which is based on the idea that there are good and bad papers, and that they may be well distinguished from each other. However, any classification process suffers from errors of first and second kind, i.e. bad contributions may be accepted and good ones may be rejected. Misclassifications occur not only due to occasional conflicts of interest. As the manuscript assessment is based on small numbers and as referee opinions vary dramatically in many cases\cite{citeulike:4085322}, the statistical validity of the individual manuscript selection process is rather questionable. In this context one should mention recent results of computer simulation indicating that a fraction of only 30\% ``rationally behaving'' referees (rejecting all papers which do not promote their own interest) is sufficient to bring down the quality of peer-review to pure random selection \cite{citeulike:7762043}. This underlines how sensitive the results of peer review are to the choice of referees.
\par
What is worse is the fact that referees often do not agree on what is a good or a bad paper (e.g. some social scientists prefer detailed models with many parameters, others prefer minimalistic models with a few parameters only, and again others do not consider mathematical models appropriate at all to understand social systems). The consistency of evaluations largely depends on the homogeneity of a scientific field and its standardization. For example, although interdisciplinary contributions are considered to be highly desirable, they typically face more difficulties to pass the referee process. This increases their publication times, reduces their average impact, and discourages many scientists from doing multi-disciplinary work.
\par
Moreover, the judgement of what is a good model or a bad one may change over time, while we often imagine that science would reveal universal, ever-lasting truths. It is even quite questionable whether one universal and consistent theory exists at all. If it {\it did} exist, it is not clear whether it would be decidable which is the right theory, given the large level of heterogeneity and randomness in human decision-making and behavior \cite{PluralisticModeling}. In fact, the success of scientific theories is often determined by herding effects and scientific fashions, and it appears to have a lot to do with social networking, not only with the brilliance or deepness of an idea. 
\par
Prominent examples of milestone papers which have been rejected are:

\begin{itemize}
\item in mathematics: Mordell's conjecture, which was later proved by Falting, winning him the Fields Medal,
\item in physics: Enrico Fermi's paper predicting the existence of neutrinos,
\item in chemistry: the paper on the Belousov-Zhabotinski reaction,
\item in economics: numerous examples are discussed in Ref. \cite{Gans}.
\end{itemize}

It is not known how many great inventions have never gained the attention of a wider audience.
The problem has probably increased with the recent policy of desk rejection by Nature, Science, PNAS, and numerous other journals. It is said that, due to the often unpredictable outcome of review procedures at high-impact journals and in order to escape the potential slow-down and copying through competitors, Nobel prize winners in high-temperature supraconductivity have chosen to publish in a low-impact journal (Zeitschrift f\"ur Physik) and to provide the correct chemical formulas only in the final page proofs. A similar strategy is said to have determined the Nobel prize winner for the discovery of quarks in elementary particle physics.
\par
It is, therefore, no wonder that the current peer-review system has been many times accused of lacking transparency, reliability, and fairness, discouraging scientific collaborations rather than encouraging them, and supporting methods and mechanisms which have been referred to as ``feudal'' \cite{WithworthPub_1,WithworthPub_2}. Only 8\% members of the Scientific Research Society agreed that `peer review works well as it is.' \cite{chubin1990peerless}.
Moreover, according to Horrobin \cite{Horrobin200151}
``A recent U.S. Supreme Court decision and an analysis of the peer review system substantiate complaints about this fundamental aspect of scientific research.'' The paper concludes that peer review ``is a non-validated charade whose processes generate results little better than does chance''. A more differentiated picture is given by a recent survey study \cite{publishingNetPR}.

For illustration, the current publication system may be compared to a funnel with a very small hole at its bottom, distilling very
parsimoniously drops of knowledge. It is questionable, however, whether creating an artificial scarcity of research space is still justified in the age of electronic publication, particularly at times where the urgency to find 
solutions to world's problems would rather appear to suggest the use of a filter or colander. 
In certain academic environments the publication system is a real bottleneck for innovation today. 

As indicated before, publication, particularly in the social and economic sciences, is a slow process which often takes two or three years, given the manuscript is accepted by the first journal.
This may well be slower than the world is changing. In case the paper is rejected and subsequently submitted
to another journal, further precious time is wasted along the way between the formulation of an idea
and its dissemination. Moreover, it is not exceptional that innovative contributions are not even considered 
for publication, particularly when they do not fit the mainstream theories and trends covered by a journal. 
\par
Also Albert Einstein had to suffer of the opposition of many 
established colleagues, before finally receiving the
deserved credit for his theory of relativity. Anecdotally, 
his response to the booklet ``100 Authors Against Einstein'' 
was: ``Why 100 authors? If I were wrong, then one would have been enough!'', but
not everybody has his standing. Many authors instead follow scientific trends, 
and mainstream economics is probably the best example of a field where researchers
have been more impressed by brilliant ideas of their colleagues than by economic
data. It is this herding effect which lets science progress through a succession of revolutionary paradigm shifts \cite{KuhnScientificRevolutions} and which made Max Planck believe that ``Science progresses funeral by funeral''. Even thought empirical evidence seems to support a somewhat more optimistic view regarding the spreading of new ideas \cite{DavidL.Hull11171978}, it is hard to deny a tendency towards large delays and substantial inertia in the adoption of new knowledge certainly exists.  

Since Max Planck's famous quote, more than 50 years have passed, but the situation has not
changed much. The reward structure of the scientific system, particularly the orientation at
high citation rates, rather tends to reinforce herding effects, while fundamental innovations
start off as minority ideas and may take a long time to be taken up, particularly if they do not happen to be promoted by big players \cite{Mazloumian}. 

Hence, the current publication system is slowing down scientific innovation, but 
not only this. 
\begin{enumerate} 
\item It discourages the replication of results due to a lack of novelty, although replicability 
is considered to be a fundamental pillar of modern science. 
\item Many journals tend not to publish papers contradicting previously published
results, since this may question the editorial process. 
If controversial contributions are not rejected by the editorial desk, they are often stopped by the referees.
\item Most journals do not publish commentaries or methodological contributions which 
could point out weaknesses of current results and questions (``grand challenges'') which
should be addressed. 
\item It is almost impossible to publish negative results, i.e. studies that did not deliver the
results one was looking for. However, describing a model or experiment that failed would avoid
similarly fruitless attempts and could help to identify successful
variants more quickly. The {\it Journal of Negative Results in Biomedicine}\cite{jnrbm} is one of the rare exceptions publishing ``unexpected, controversial, provocative and/or negative'' results.

\end{enumerate}

All these circumstances slow down science and do not use resources (funding and manpower)
efficiently. In particular, it can take a long time until incorrect or useless results are revealed, and generations of 
PhD students may be wasted on paths which eventually turn out to be dead ends. If the publication system could be changed such that important results would be confirmed more quickly, authors would be more careful to make sure
their results are replicable. Also, scientific fraud would be discovered much earlier.

In summary, the way in which science is organized today does not appear to promote the efficient collaborative
solution of problems that humanity is facing. First of all, there are only few methodological contributions
which identify and raise the crucial questions. Second of all, the system is more competitive than what seems
to be good for collaboration. The research and publication process is slow and wastes resources.
Multi-disciplinary contributions, although urgently needed, often do not make it into high-impact journals.
In fact, for interdisciplinary work and socio-economic contributions, the current publication system lacks
established science journals beyond Nature, Science and PNAS with a rewarding impact factor. 
Moreover, most papers that {\it do} successfully pass the review process are never read or cited 
\cite{RobinsonPub}. According to Meho \cite{citeulike:3039453}, ``Only 50\% of (accepted) peer-reviewed articles are ever read by someone other than the authors and the reviewers. Furthermore, 90\% of articles are never cited.'' 
\par
Heterodox contributions have particular difficulties to be published and noticed. Furthermore, scientific controversies tend to be discouraged, although they would probably be very stimulating. Finally, a curiosity of the scientific production system is that, in contrast to many other creative areas, authors write (and review) for
free, while their institutions pay a high price to be able to read contributions of others. Alternatively, if they want to make their contributions freely available to everyone (including smaller universities and less developed countries), they have to pay 
a considerable price for open-access publication.

\subsubsection{Emerging trends in scientific publication}

Recently, an intense discussion about the future of scientific publishing has set in. This is reflected by a number of contributions and events \cite{WithworthPub_1,WithworthPub_2,RobinsonPub,Allesina_Bidding,ScienceOnline,FacebookScholars,GeniusIndex,nielson09,Science2.0} and by the fact that Europe has lately been funding projects addressing this issue \cite{LiquidPub,Openaire,Visioneer_site}. 

This has also created a number of changes in the publication landscape and the behavior of scientists. For example, preprints of publications are now often uploaded to electronic archives before their acceptance and publication in a journal. The most well-known example is probably arXiv.org \cite{ArXiv}. Some communities, such as in high-energy physics, seem to have even replaced journal publications by archive publications to a certain extent. Archives are also more and more used in the social sciences (e.g. SSRN \cite{SSRN}). 
\par
In an analysis of the advantages of open access publishing in High Energy Physics (HEP), Gentil-Beccot {\em et al.} \cite{citeulike:5034439}
summarize: 
\begin{itemize}
\item ``Submission of articles to an Open Access subject repository, arXiv, yields a citation advantage of a factor of five.''
\item The citation advantage of articles appearing in a repository is connected to their dissemination prior to publication, 20\% of citations of HEP articles over a two-year period occur before publication.
\item HEP scientists are between four and eight times more likely to download an article in its preprint form from arXiv rather than its final published version on a journal web site.''
\end{itemize}
As a consequence, they conclude:
\begin{itemize}
\item ``There is an immense advantage for individual authors, and for the discipline as a whole, in free and immediate circulation of ideas, resulting in a faster scientific discourse.''
\item ``Peer-reviewed journals have lost their role as a means of scientific discourse, which has effectively moved to the discipline repository.''
\end{itemize}
However, what worked in the well-networked HEP community does not seem to be directly transferable to other disciplines. In areas different from particle- and astro-physics, arXiv does by far not cover 100\% of published papers, and the fraction of manuscripts uploaded there is apparently not converging to full coverage \cite{arxiv_mith}. Also the establishment and spreading of recent public access journals (such as the Public Library of Science, PLoS) and of public access options of classical journals suggests that there is still a considerable interest in journal publications. This may relate to the promise of publishers to keep scientific results accessible over long time periods (over which old file formats may disappear and new ones may become temporary standards). Moreover, journals seem to play a role in terms of dissemination (marketing). 
\par

\subsection{Science 2.0: A new open framework for scientific publication}\label{Science2}

While facing the intricate challenges of our century, such as AIDS, cancer, climate change, the 
financial crisis, poverty, etc., it is our responsibility to make sure that available knowledge
is recognized and used, and that new progress is made efficiently.
In order to accomplish this, we need to create suitable ICT systems to produce, share,
filter, combine and present scientific discoveries. This is only slowly
happening in science. We are still wasting time due to the wide application of outdated
concepts, technologies, and incentives. In the following, we will sketch a concept that shows how current journals may be developed further,
combining the advantages of classical journals and archives, and providing community-based quality
selection mechanisms to discover the pearls among a large quantity of scientific
contributions. In fact, suitable filtering and discovery techniques may replace the practise of excessive review and revision procedures (which are very time consuming for scientists today), cutting them back to a reasonable level. 
A journal of the envisioned kind may be multi-disciplinary in nature and imagined as outlined in the following subsections.

\subsubsection{Archiving}

The basis of the proposed journal of the future would be an archive, where manuscripts could
be uploaded by authenticated users. To save hardware and maintenance costs,
the archive could be based on a decentralized platform such as a peer-to-peer
system. If several preprint platforms are used in parallel, interoperability would be a 
desirable feature. A search engine for scientific papers would then pull
the information from all platforms together and integrate them in one
portal. It would be able to present the search results in a unified way.

As soon as a manuscript is uploaded to any of these archives, it would
become available for public download world-wide. As orientation for users,
a usage statistics could be provided (number of views, number of
downloads, etc.), reflecting the popularity of the paper. However, to avoid intensifying the 
Matthew effect \cite{citeulike:1040330} beyond what seems to be useful, it would make sense
to randomize the order of display to a certain degree.

According to the storage mechanism of the decentralized archive, papers that have not been accessed for a long time would lose visibility, while frequently accessed papers would gain visibility (and could be downloaded more quickly). 

\subsubsection{Peer Review}\label{Review}

Today's peer review system tends to overload referees with papers they may not even
be interested in reading (remember that most published papers are never downloaded or cited, see above).
Therefore, in future not all manuscripts (preprints) should be reviewed anymore. 
However, in order to promote scientific quality,
specialized editors would select certain archive contributions (preprints) for an anonymous peer review procedure.
Reviews may also be submitted by readers. In this way, a top-down selection would be combined with a 
bottom-up (``grass-roots'') procedure, given some mechanisms to ensure quality control. 

Referees would be asked to make critical contributions which
help the authors to improve the quality of the manuscript in a revision round. Papers which had at least one positive report out of three would be considered for revision, given the positive referees have 
enough reputation points (see Sec. \ref{Reputation}). The anonymous referees would write comments on the revised manuscript, which are published in an electronic journal together with the paper and with replies of the authors to these comments (given there  
was a positive editorial board decision based on the referee comments, author replies and revisions, which would be the standard case). Moreover, journal papers would get an initial rating by the referees and editorial board. In other words, a quality improvement mechanism would apply, but the publication of potentially valuable scientific contributions would not be suppressed. 

The new assessment procedures would make sure that scientific journal publications 
would have high standards and that the scientific debate would be stimulated. 
Referees would identify the weak points of a manuscript, while authors would have full responsibility for their quality and a fair chance to refute criticism. Rejected contributions would remain in the archive and be publicly accessible for a long time. All relevant agents in this process (i.e. authors, referees, and editors) would be rated (see below), i.e. they could gain or lose reputation depending on the quality of their work. The risk of negative ratings, comments, or replies would encourage everyone to do their job well. 

\subsubsection{Journals}\label{Journals}

Based on the wide range of disciplinary and multi-disciplinary contributions undergoing the peer-review
procedure, different kinds of disciplinary journals should be created to target certain readerships. Multi-disciplinary contributions could appear in different disciplinary journals. These journals would usually appear in print (printed information has probably still the longest survival time over a period of several hundred years). A small subset of contributions would additionally be highlighted by ``best-of editions'' or ``editor's choices''. These would serve similar functions as today's letter magazines, trying to promote the rapid communication of particularly important results, but they would not suffer from space limitations. 

Journals would also stimulate scientific debates and organize contributions according to different paper categories (see Sec. \ref{categories}). 
Besides supporting a wider scope of contributions, journals could be improved
in a number of ways, building on opportunities offered by what is called the Web2.0. 
A list of various Web2.0 tools and platforms for science is provided in the Appendix \ref{web2.0}.
These technologies could, for example, be used for an individually customized information filtering and retrieval system, and to support scientific convergence by interaction rather than selection. In fact, repeated social interaction is known to support the convergence of ideas in a natural way \cite{LorenzRauhut}. A proposal of how a journal could be made up in the era of Web2.0 is presented in the next section.          

\subsubsection{Web article presentation}\label{WebArticlePresentation}

Journals should present each reviewed paper on a dedicated
Web2.0-like portal, which would be a Web environment 
stimulating the scientific discussion of scientific subjects.
Comments and replies relevant to the topic should be supported.
Comments would usually not be anonymous, but in certain cases
this may be acceptable. To avoid inappropriate content, anonymous
comments would be monitored by moderators chosen by the editorial board. Discussion should be stimulated by implementing suitable incentives schemes for contributions \cite{ScienceMonologue,Goetzsche} (for example, the reputation system outlined in Sec. \ref{Reputation}).
 
One crucial feature of the Web2.0 platform would be to support the
rating of articles, authors, referees, and editors. These ratings could be
carried out by registered users and they would be visible to everyone. 

The Web page would link a publication to related contents such as

\begin{itemize}

\item older versions of the same article (visual diff tools could help
  to highlight changes between different versions),

\item related multimedia files and other supplementary materials,

\item cited articles and, even more importantly, missing citations as determined
by a suitable algorithm,

\item the dataset used,

\item related materials as listed in Sec. \ref{categories}.

\end{itemize}

The Web2.0 platform should also be endowed with a recommender system, indicating in particular
relevant contributions from other fields (see Sec. \ref{RecSys}). It should
furthermore provide a forum for the discussion of work in progress (something like question and
answer channels, which could be run by subject-specific user groups).

\subsubsection{Paper categories}\label{categories}

In the age of electronic publication, the dissemination of information has become much
cheaper, and there is no limitation of pages. Besides, it would be easy to
create a {\it network} of interrelated contributions. When doing so, a number of different 
categories should be distinguished:
\begin{itemize}
\item Original research papers, reporting results of recent studies (which is the only or at least the
main category of most journals),
\item proposals and methodological papers, identifying research needs and elaborating important questions, e.g. scientific grand challenges,
\item review papers, summarizing the progress in a field,
\item opinion contributions, allowing for subjective judgements,
\item summaries, such as conference reports or book reviews, insights into other fields, etc.,
\item replications reporting an independent confirmation of a result, preferably with another method,
\item contradictions which report results that are inconsistent with previously reported findings,
\item negative results reporting unexpected failures one may learn from,
\item errata (corrections of previously published results, e.g. mistakes that were made in calculations),
\item controversies which would promote a critical dialogue between different points of views on a certain subject,
\item blogs and podcasts, highlighting particularly relevant advances,
\item interviews, asking for points of views on certain subjects and revealing implications of certain findings, e.g. for society, technology, and economy,
\item comments and replies (discussions) which reflect on any of the other categories.
\end{itemize}
All these contributions, linked in a network-like structure, could be supported by a set of digital libraries, including videos and other resources, and offering real meeting points for discussing and possibly video recording arguments and controversies. However, it should not be forgotten that also the classical structure of papers itself can be significantly improved. For example, authors should be requested to formulate in special (sub)sections as precisely as possible
\begin{enumerate}
\item the research question (puzzle, challenge, ``mystery'') addressed,
\item the research methodology/approach,
\item the underlying assumptions,
\item the current evidence,
\item implications or predictions allowing to assess the explanatory power,
\item the expected range of validity or limitations of the approach.
\end{enumerate}
                      
\subsubsection{Editorial board}\label{Board}

Editorial boards have a fundamental role in stimulating high-level scientific exchange. 
They should be composed of people who have had impact
and earned international reputation already. In order to concretely
define who is entitled to sit on a board, minimum scientific requirements 
should be formulated. For example, one may start with scientists who have
published in leading journals and/or have had several publications
with 100+ citations. Additional editorial board members may be elected
by the editorial board, if this is necessary to fill gaps in the coverage of certain 
areas. In the long run, however, the editorial board would be composed of the
scientists who reached the highest reputation points (see Sec. \ref{Reputation}). Finally, turnover policies should be implemented, which regularly exchange editorial board members in order to avoid their overload and to counteract the emergence of systematic biases in favor or against certain fields, approaches, or authors.

\subsubsection{Rating}\label{Rating}

Journal contributions would be rated on a multi-dimensional 
scale, based on criteria such as readability, importance, novelty, controversy, etc. Registered
users could rate a contribution once and only once, and rectifications
should be possible. Users would be endowed with a fixed amount of rating points per month. 

Regarding the actual rating widget, special solutions should be
implemented that allow to visualize multiple dimensions at the same time.
For example, ratings in different categories could be displayed in different
customized ways, such as
\begin{itemize}
\item bar diagrams with bars in different shapes and colors, 
\item star diagrams, or
\item facial diagrams.
\end{itemize}
Besides the colors, also their saturation and luminosity values would
be varied. Initial ratings would be in light colors, but the rating symbols
would become more intense, the more ratings are made, while a large
variability of ratings would reduce the color intensity. 
\par
Depending on their focus, each journal could select their own rating criteria and the weight they
are giving to them. For example, they could weight certain dimensions like novelty or
controversy more than others. 

\subsubsection{Reputation system}\label{Reputation}

Besides journal contributions, also raters would be rated and evaluated. That is, authors, referees, and editors would 
earn a reputation, which would determine the weight they have in the determination of average
ratings. Higher reputation would also imply certain user benefits (see Sec. \ref{Incentives}). 
Therefore, raters should be concerned about their reputation.

Reputation and reputation systems have been the object of scientific study already for some time \cite{conte2002reputation,sabater2005review,farmer2010building,farmer2010building,paolucci-electronic,brabazon2006google,grimaldo}. For most collaboration systems, reputation and its management is the key point that decides over success or failure. Challenges to address include the identification of communities \cite{2010PhR...486...75F} and of collusion or defaming \cite{352889}
(see Sec. 7 in \cite{VisioneerCrisis} for some related proposals).

Reputation could be determined, for example, from the average ratings
of the contributions of a rater. Unusual rating patterns (as compared
to other users) would be identified by algorithms, to reveal possible misuse
of personal reputation in rating activities. (In case of disagreements with general
opinion trends, it would be appropriate to make a comment.) Attempts to manipulate 
the rating system would be sanctioned, e.g. by setting the weight to zero for some time.

\subsubsection{Recommender system}\label{RecSys}

Another challenge besides the design of a manipulation-resistant
reputation system is the creation of a privacy-respecting recommender system. 
Both issues have been addressed in another White Paper \cite{VisioneerCrisis}. 

The Web2.0 science journal could offer their users to customize their own rating
system. Consequently, they may create their own indices by setting personal weights
determining the ranking of papers (e.g. they may overweight novelty or controversy).
Of course, they would also specify their field of interest.

The system would recommend contributions based on keywords, title and abstract (tagging concept), or based on correlations in download patterns (Amazon concept). Alternatively, a user may choose to get recommendations on author network analysis or citation analysis, or on recommendations of certain raters, or a surprise mechanism.
Users may also be alerted when new manuscripts of interest (e.g. in certain subject areas or by specific authors) appear.
The recommender system may also analyze the impact of papers or authors, or determine emerging fields.

\subsubsection{Incentives}\label{Incentives}

So far, a number of journals have made experiments with several of the above features, but the success has
been limited. The main reason seems to be that user participation in rating and commenting tends to be low.
In order to change this, users need to have incentives to contribute.

For example, the Web2.0 journal could foresee different kinds and levels of access to the publication, 
information and recommender system. Depending on the amount of user contributions and their
quality (as determined by the rating system), users may have earlier or later access to newly published
articles, or they may be able to use certain functions of the recommender system or not.

The number of rating points per months could be coupled with the number and quality of contributions as well.
Furthermore, the reputation of contributors would determine their weight as referees. Contributors with high
reputation would also qualify for bottom-up review (see Sec. \ref{Review}). 

Contributors with particularly high reputation would qualify themselves for the editorial board (see Sec. \ref{Board}). To provide incentives to work as editorial board member, these may be allowed to select number (e.g. three) of their papers per year for the ``best of selection'' (see Sec. \ref{Journals}). Furthermore, they may select (``sponsor'') a limited number of papers of other authors for it. The other contributions in the printed ``best of'' volumes would be determined according to the ratings of the scientific community.

\section{ICT solutions to increase the awareness of new emerging
  trends}

Ideas and successful innovations often start off in a minority position \cite{Mazloumian}, and they  
are hard to notice in an information-rich environment. They require targeted support in order to
flourish. Therefore, suitable tools are needed to identify emerging fields, rising stars, and
natural scientific alliances early on. 

IBM's shortsightedness in foreseeing the potential of personal computers is 
just one example for the difficulty to determine the potential of innovations. 
It is well reflected by the famous 1943 quote by Thomas J. Watson (Chairman of the Board of International
Business Machines - IBM): ``I think there is a world market for about
five computers.'' Even more sensational was Xerox's donation of the
mouse concept as we know it today to Apple during a visit of Steve Jobs to the Xerox PARC research center. This was
1979, and as it is well-known that it significantly contributed to the great
success of Apple computers. Also, the widespread use of
text messaging on mobile phones was not at all anticipated.

Fortunately, these technologies eventually found their way,
but many potentially useful innovations never did, and this is true
for both industry and science. We therefore need tools to systematically
discover the most innovative ideas. 

\subsection{Classical scientific impact analysis}\label{Classical}

In the landmark contribution ``On the electrodynamics of moving bodies'', in which Albert Einstein elaborated
his theory of special relativity in 1905, readers will be surprised to find no references to other scientific publications.
Today, the number of citations is often considered to be the gold standard for rating a scientists' value. 

Knowing of the benefits and limitations of citation analyses, there are a number of different indices that try to
identify good scientific contributions, authors, and journals, as is reflected by the following incomplete list \cite{Pop,EigenFactorExplained,HalfLifeExplained}: 

\begin{itemize}

\item \textbf{Hirsch's $h$-index.} The most famous index for measuring scientists quality
is named after its inventor Jorge Hirsch \cite{Hirsch15112005}.
It is defined as the maximum integer $n$ such that there are $n$ papers
which received at least $n$ citations each. 

\item \textbf{Egghe's $g$-index}
Proposed by Leo Egghe \cite{G-Index}, it aims to improve on the $h$-index by giving more weight to highly-cited articles. It is defined as follows:
Given a set of articles ranked in decreasing order of the number of citations received, the $g$-index is the largest number such that the top $g$ articles received on average at least $g$ citations.

\item \textbf{Zhang's $e$-index}

The $e$-index as proposed by Chun-Ting Zhang \cite{citeulike:4486103} tries to differentiate between scientists with similar $h$-indices, but different citation patterns. It is the square root of the surplus of citations in the $h$-set beyond the theoretical minimum required to obtain an $h$-index of $h$.

\item \textbf{Contemporary $h$-index}

Proposed by Antonis Sidiropoulos, Dimitrios Katsaros, and Yannis Manolopoulos \cite{citeulike:846275}, it aims to improve the $h$-index by putting more weight on recent articles, thus rewarding scientists who maintain a steady level of activity.

\item \textbf{Age-weighted citation rate (AWCR) and AW-index.}

The AWCR \cite{jin2007ar} measures the average number of citations to an entire body of work, adjusted for the age of each individual paper.

\item \textbf{Individual $h$-index}

Proposed by Pablo D. Batista, Monica G. Campiteli and O. Kinouchi \cite{citeulike:5154380}, it compares researchers with different scientific interests. It divides the standard $h$-index by the average number of authors in the articles that contribute to the $h$-index, in order to correct for the beneficial effects of co-authorship.

\item \textbf{Multi-authored $h$-index}

A multi-authored version of the $h$-index by M. Schreiber \cite{citeulike:2652200} uses fractional paper counts instead of reduced citation counts to consider the shared authorship of papers. It determines the multi-authored $h_m$ index based on the resulting effective rank of the papers, using undiluted citation counts. 

\item \textbf{$h_b$-index and $m$-number.} 

The $h_b$-index \cite{citeulike:625569} has been developed as another extension of the $h$-index. It applies to scientific topics instead of to individual scientists. Assuming that $h_b$ increases linearly with the number of years $n$ from the first published paper in a given topic, the $h_b$ index can be defined as $h_b = nm$, where $m$ is the gradient, which varies from topic to topic. Large values of $m$ and $h_b$ denote hot-topics.

\item \textbf{Journal Impact Factor (JIF)}

The Journal Impact Factor has been invented by Eugene Garfield in 1975 \cite{citeulike:584493}. It is a proxy for the relative importance of a journal within its field. Given a reference year, say 2009, it is defined as the total number of citations received in 2009 by papers published in the journal in the previous 2 years, divided by the number of these papers. The journal impact factors are calculated yearly for those journals that are indexed in Thomson Reuter's Journal Citation Reports.

\item \textbf{Immediacy index}

It reflects the average number of citations that articles in a given journal receive during the year they are published. It is part of the yearly Journal Citation Report, calculated by ISI \cite{ISIKnowledge}.

\item \textbf{Cited half-life}

The Cited Half-life is the median age of articles that were cited in the Journal Citation Reports each year. For example, if a journal's half-life in 2005 is 5, it means that the citations from 2001-2005 are half of all the citations of that journal in 2005. The other half of the citations precede 2001. 

\item \textbf{Aggregate impact factor}

The aggregate impact factor is determined from the number of citations of all journals in a subject category and the number of articles in all journals belonging to that subject category. 

\item \textbf{\textit{Eigenfactor\texttrademark} Score}

The Eigenfactor Score calculation is based on the number of times articles from the journal published in the past five years have been cited in the Journal Citations Report year. It also takes into account the impact factors of journals and eliminates the effects of journal self-citation. 

\item \textbf{\textit{Article Influence\texttrademark} Score}

The Article Influence shows the average influence of a journal's articles over the first five years after publication. It is calculated by dividing a journal’s Eigenfactor Score by the number of articles in the journal, normalized as a fraction of all articles in all publications. 

\item \textbf{Co-citation index}

Two documents are said to be co-cited if they appear simultaneously in the reference list of a third document. The co-citation frequency is defined as the frequency with which two documents are cited together. The co-citation index has been first proposed by Henry Small \cite{citeulike:2498154}.

\end{itemize}

\subsection{New indices to discover innovations}\label{New}

The indices mentioned in the previous section can certainly be helpful to determine promising scientific contributions, but they also have a number of weaknesses. For example, they are not rewarding scientific contributions that are not separately cited (such as data sets or computer animations or many of the contributions called for in Sec. \ref{categories}). Moreover, scientific impact is still largely measured in a journal-centric way, while it should be measured directly where it matters, i.e. on the articles level \cite{ImpactWhereMatters}. Therefore, in an attempt to extend classical bibiometric measures, a new discipline called Scientometrics 2.0 is trying to mine Web 2.0 sources, such as clickstreams \cite{10.1371/journal.pone.0004803}, downloads, news, tweets, Diggs and blog entries, looking for signals of scholarly impact \cite{Priem10}. Citations in traditional journals may take years to accumulate \cite{Mazloumian}, while on the Web the community response can be measured almost immediately. Successful further steps in this direction could permit to create a quasi-real-time monitoring of cutting-edge scientific innovation across disciplines.

Another dissatisfactory point of the current way of measuring impact is the circumstance that the number of citations does not reflect the relative scientific importance of a contribution well, as some fields are small and others are large. One consequence of the orientation at citation rates is, therefore, that scientists are pulled into highly cited fields. Such herding effects cause that people turn their attention away from other important research fields, particularly from difficult ones with low publication and citation rates. 

One would therefore need to have indices allowing one to compare contributions and scientists from different disciplines, and to judge multi-disciplinary publications in a fair way. A first step in this direction has been recently made \cite{Radicchi11112008}: The $h_f$-index
proposed by Radicchi, Fortunato and Castellano aims at reducing the strong field dependence of the $h$-index due to different sizes of scientific fields and their heterogeneous publication rates.
It is computed like the $h$-index, but after scaling both the number of citations and the rank of the papers by suitable constants depending on the discipline. Unfortunately, these constants are currently not available for all fields.

In addition, it would be important to compare the potential of contributions by junior scientists with those by senior scientists. Consequently, suitable indices will have to be developed for this. Recent analyses give hints how this could be done \cite{Mazloumian}.

A further relevant question is, how to translate performances measured on multiple scales into one single indicator (i.e. a rank). This is traditionally done by weighting each criterion with a certain factor. Such an approach, however, promotes average performance rather than excellence, as the latter is typically characterized by extreme values on one or a few rating scales, but not in all of them. In order to give everyone a fair chance and to reward excellence in specific areas, one needs to introduce new metrics capable of spotting out individual talents. Appendix \ref{pleval} suggests two promising ways of doing this. 

Finally, it seems to be advisable to complement citation analysis with reputation analysis, as outlined in Sec. \ref{Reputation} and in Ref. \cite{VisioneerCrisis}. In order to determine the potential of innovations, it is also necessary to separate the effects of institutional settings (such as better equipment or better networking, which can largely accelerate the dissemination of scientific work) from the quality of
contributions (which reflect individual talent). Spotting the right
talents and the right institutions is quite important. Moreover, it
would be desireable to quantify conditions of scientific success, such
as a multi-disciplinary collaboration culture. Questions like these
are recently being addressed by research fields like
Scientometrics \cite{citeulike:6463676,citeulike:4680670}, or, more recently, Science
of Science \cite{KatyBorner,citeulike:4908809}. Both fields are largely overlapping,
but the latter focuses more on the financial, social, geographical and institutional factors contributing to scientific success.

\section{Tools to enhance Europe's innovation potential}

\subsection{Why and how to free up scientists' time for research}

A recent Europe-wide initiative, called Trust Researchers \cite{TrustResearchers} has pointed out an urgent need to reduce the administrative overload of scientists and find better ways to fund research. In fact, great scientific talent is extremely rare, and consuming time of these talents for anything else than science is an extreme waste of resources. It causes that less high-level research can be performed, a problem that can only partly be compensated for by employing additional (less talented) scientists. Consequently, it is imperative to protect scientists from overusing their most precious and scarce resource, which is \textit{time}.
\par
In spite of this, over the last decades, scientist found less and less time for research, as they had to pay more attention to teaching, to a growing number of students (with, consequently, less talent on average), on grant applications, managing projects, preparing reports, presenting at project meetings, on public dissemination through the media, etc. This large variety of activities is extremely distracting and makes it impossible to concentrate on one task over extended time periods, as it is typically required for breakthroughs. However, fractionalization of time into small pieces is just one problem. A fact known from logistics is that it takes a massively longer time to complete tasks, when one operates too close to maximum capacity (time reserves), and most scientists already work much longer than the time they are paid for (often 60 hours per week or more). 
\par
The fact that scientists are performing many additional tasks without extra payment (typically) has caused a ``tragedy of the commons'' \cite{citeulike:435835} in the scientific system. While evaluations were focused originally on evaluations for tenure and search committees, they are now made for a steadily growing number of papers and project proposals, for students, even for conference applications. This cancerous spreading of evaluation load is a serious waste of resources. While the classical idea of Humboldt's university concept foresees 50\% research and 50\% teaching, administrative loads can easily exceed 80\% today. Estimates regarding the time spent in EU projects on administrative tasks (including proposal writing, coordination, meetings, reporting, evaluation, financial accounting, presentation, dissemination) reach from 40\% to 75\%. It is obvious that tax payers' money could be used more efficiently, if better funding schemes were available. 
\par
Given the scientific performance indices that are available today or under development, one could move from funding of promised research results (proposals) towards refunding for research results obtained. Scientists would then focus on research and the publication of their results. Proposals, intermediate and final reports would not be necessary anymore. Instead, publications would be evaluated. This would free up time for research and publications, which are anyway subject to a quality evaluation process. In fact, Universities in the Netherlands and in China, for example, grant money for publications.  This funding principle could be largely extended (see the paragraph on ``incentive-based crowd sourcing'' in the next section \ref{howto_IA}). Research proposals would then only be needed for special investments (e.g. expensive laboratories) which cannot be covered by overheads.  
\par
However, scientists are not only burdened by administrative and managerial tasks. 
Finding relevant information for their studies becomes a
more and more inefficient process due to the world information overload 
that is sometimes called ``data deluge'' \cite{VisioneerCrisis}.
For example, in 2008 there were about 47,000 papers and more than 350,000 datasets containing useful
information about the p53 protein which regulates the cell cycle and that
could prevent the development of cancer \cite{NeurocommonsIntro}. This
means that, at present, there is no way for a single human being to
browse through and get all the possible knowledge out of such a vast
literature. Yet the problem of finding a definitive cure for cancer
remains open. 
\par
In other words, researchers must waste a considerable amount of their
time in mining vast scientific corpora with inefficient techniques,
looking for the most significant contributions. Obviously, finding
related work in complementary disciplines is even harder. 
Therefore, scientists need the support of specific searching, archiving, sharing and
discovery tools. Conditions should be provided, in which they can devote as much of
their time as possible to the creation of quality. The ``Innovation Accelerator'' 
or ``Knowledge Accelerator'' sketched in the following, could create such conditions.
\par
The Innovation Accelerator is an integrated ICT-based platform aimed
at fostering the creation and sharing of scientific excellence by
reducing all unnecessary friction of today's scientific knowledge production and 
dissemination. It will help business people, politicians and scientists to find the
best experts for a project, ease the communication in large-scale projects and support
their flexible coordination, co-creation, and quality
assessment. New trends will be discovered earlier on, allowing the
investment into emerging trends and technologies. 
\par
The Innovation Accelerator requires the provision of new tools to
\begin{itemize}
\item support a community-specific definition of scientific quality,
\item easily setup and manage large-scale scientific collaborations,
\item allow efficient scientific co-creation,
\item allow efficient many-to-many communication,
\item promote schemes for a fair distribution of public funding based on scientific merits.
\end{itemize}
The innovation accelerator is expected to trigger many positive externalities, e.g. 
\begin{itemize}
\item to increase interactions among scientists,
\item to stimulate scientific debates,
\item to promote the exchange between different scientific communities,
\item to provide better chances for scientific innovations and heterodox research approaches,
\item to support all steps in the scientific production process. 
\end{itemize}

Some of the principles that will be required to make such a system
work are a balance of power (symmetry), transparency, feedback, sanctioning of misuse, and
ownership of, responsibility for and control of results of creative activity. 
In the next section we will describe how such a Web 2.0-like, distributed platform
could look like, and what features it should have.

\subsection{How to create an Innovation Accelerator (IA)}\label{howto_IA}

Quite recently, it has been impressively demonstrated how powerful massive
collaboration can be used to solve complex problems, for example, in mathematics \cite{citeulike:5943790}.
The Innovation Accelerator (IA) is envisioned to be a tool to support such
large-scale creative collaborations. It can be imagined as a distributed
internet-based platform, implementing the trinity of zero-install, 
ubiquitous access and rich and intuitive UI (User Interface) \cite{Web2.0Fundamentals}. 

The IA framework could be realized through the
use of standardized building blocks that are explicitly created to communicate
with a large information network infrastructure, acting as
backbone for sharing data across multiple communities. Each building
block would represent an independent entity, offering a well-defined service
hidden behind a standard interface. Finally, an intuitive
administration panel would permit each community to create a customized
IA tailored to their needs, combining the desired
blocks in a simple-to-use, but powerful tool.

In order to promote innovation rather than obstructing it, the use of 
the IA should be free or at least affordable for academic 
institutions all over Europe. For example, the IA architecture could be open-source 
in order to reach the openness and dynamics ensuring that it is well functioning 
and widely used. The IA architecture would include the following modules:

\begin{itemize}
\item A \textbf{forward-looking resource manager} optimizing the use of resources (money, space, staff, etc.). For   example, it should be able to suggest different options
  how project money at an institute (potentially coming from different funding sources with different spending restrictions)
  would be best spent, considering plans and constraints. 
  
\item A \textbf{project and team calendar} would support a coordinated project schedule and 
  send reminders or alerts, when appropriate.

\item An \textbf{intelligent career manager} would try to 
  match job openings and the best experts on a European scale. For 
  example, required competencies could be searched against a tag
  database or against similarity scores in recent publications.

  Interested researchers would create a list of institutions where
  they would like to work, and they would be automatically notified of
  the next available positions. On the other hand,
  institutions could tune the number of desired applications by
  requiring to pass a certain reputation barrier or test before
  applying. 
  
  The same system could determine an appropriate salary range, based on the 
  respective set of skills, publications, and other factors. 
  
\item A \textbf{social networking module} 
 would provide standardized building blocks for the creation of project websites 
 and discussion groups (e.g. workpackage-specific ones). It
 will offer standard interfaces to organize and combine the
  other components of the IA in a fully customized way.

  Clicking on names (or pictures) of the mutually entangled social networks would 
  bring up their most recent contact information, and to alleviate personal
  contacts, would indicate at which conferences one could meet the
  person (given his or her permission). One could also directly e-mail or skype
  the person or would get an information when the person is reachable
  (given this information is made visible).

\item A \textbf{many-to-many communication system} will allow one to
  manage complex messaging patterns within a simple and intuitive
  interface. For example, users will be able to activate specific (e.g. workpackage-related)
  mailing groups within a few clicks, tuning parameters such as adding or removing members,
  enabling or disabling the reply-all function, blind-carbon copies,
  hiding the email addresses of recipients, but showing their names, etc.
  It should manage groups easily via the social networking module and automatically
  resolve address duplication and address updating issues. 
  The messaging should support the inclusion of
  crowd sourcing widgets such as polls, doodles and maps. 

  Additional security and reliability protocols could be enabled directly
  from the same interface, and encryption and/or digital signatures of messages should be easily possible. Other procedures such as public key
  retrieval, handshake phases, etc. should be automatically handled by
  the system.

  Sanctioning mechanisms against spammers and other abusers should
  also be implemented. 
  
   \item A \textbf{virtual conference module} would allow one to set up Web seminars (``webinars'') or
   Second-Life-like environments for virtual group meetings, 
   thereby reducing the need for travelling. Mechanisms for moderating,
   assigning turns, and reserving the next speech should permit every
   participant to express opinions in an orderly fashion. 

   Participants of the virtual meeting would be supported by a series
   of virtual gadget. For example, electronic documents such as
   articles, images, videos or maps should be easily and immediately
   accessible. Moreover, virtual dashboards would capture in real-time
   all relevant information of the meeting, store it in an encrypted file and
   send out a link and decryption information to authorized recipients who want to have access to the recordings.

\item An \textbf{incentive-based crowd sourcing system} would collect ideas
how to address the most important scientific challenges. These would previously
be elaborated at ``Hilbert workshops'', at which scientists gather {\it not}
to present their results to each other, as usual, but to identify open problems. 
The resulting set of questions would be published on-line, and there would prizes to 
reward the best solutions. Prizes could be diverse, ranging from a ``medal'' or monetary prize 
up to research grants or academic positions.
The selection and formulation of the grand challenges should
make sure that practically relevant, goal-driven, and multi-disciplinary research 
would be particularly stimulated. Such an approach has been very successfully applied by 
Innocentive \cite{Innocentive} or in the DARPA's balloon-finding challenge \cite{DarpaBaloon}. It is also being used in certain software development communities, for example in discovering and documenting bugs in new releases of Internet browsers \cite{GoogleBounty}. A comprehensive collection of prizes issued to stimulate innovation can be found in Ref. \cite{kei_prizes_innovation}. Of course, less fundamental challenges could be posted in separate subject- or
community-based user forums and worked out in a similar way, according to
customized settings. 

\item A \textbf{public dashboard} will allow people to announce the
  current subject of their study. This serves to stimulate collaborations and 
  to avoid too much multiplicity in attacking scientific challenges. 
  For example, scientists working on a certain challenge will
  be displayed next to it, and additional information such as
  the solution approach may be announced as well. They could also
  look for partners with competencies they are lacking themselves
  (see the ``intelligent career manager'', the ``information discovery and
  retrieval system'' and the ``reputation system'' for details). Ideally, the system
  should foster both collaboration and competition to the right extent
  (``coopetition''). 

\item A \textbf{decentralized co-creation system} would
  allow researchers from all over the world to actively and
  efficiently take part in large-scale projects. Commonly produced
  documents would be stored within a versioning system which would keep
  track of all the changes and highlight them, with the possibility to easily revert or merge
  them by semi-automatically resolving conflicts between diverged versions.

 Ideally, project participants would be able to work on different parts
 of the document in parallel.  Mechanisms to assign or reserve portions of the document 
 for a given amount of time to certain editors should be implemented (through suitable access
 right management). Moreover, the system should automatically invite people to work on 
 certain parts, and it should highlight who is currently editing them. 
 
  Document versions and their sections and paragraphs could be rated 
  \begin{itemize}
  \item[(i)] by the authors themselves,
  \item[(ii)] by invited reviewers or
  \item[(iii)] by public audience (according to
  what has been decided by the authors). 
  \end{itemize}

  Different roles for commenting
  and rating could be easily implemented here.

  When the quality of the ratings reaches a certain level, which means
  that the majority of authors has found an agreement, that part or version of
  the document is frozen and eventually submitted for publication. Authors
  could continue working on other parts or on new versions of the document.

The system would record the activity of each author, and should be able to visualize their contributions in the joint document. Author contributions may also be summarized in words or by statistics at the end of the document. This would help to clearly identify who did what, which can usually only be guessed from the order of authors names (a situation that is particularly dissatisfactory when a publication is co-authored by many scientists). In fact, journals like Nature and Science have started to explicitly state Author Contributions.
  
\item A \textbf{quick annotation system} would store and retrieve
  important notes for the future, and it would allow one to easily import (``pull in'') graphics, 
  statistics, Web links, videos, scientific references, and other relevant information. 
 
\item A \textbf{semi-automated reporting system} should take care of 
  reporting duties. This requires that the information (about
  publications, projects, and presentations, etc.) would be
  standardized, searchable through a single tool, and entered only
  once, while the resulting output could be formatted automatically in
  individually customized ways (such as a publication list, for
  example, or a CV, or annual report).

\item An \textbf{information discovery and retrieval system} would
  consist of a rich user interface allowing to easily and simultaneously
  query multiple databases and archives. When searching information,
  people could choose between different options. Information of
  interest would be determined based on information such as time-window of generation, 
  author name,\footnote{Advanced name disambiguation
    algorithms would be automatically invoked whenever required.}
    tags (key words), ratings, author reputation, number of comments, number of downloads, etc. 
    (see Sec. \ref{RecSys}). Advanced search criteria could be: user-defined compound indices,
  correlations in download patterns, author network analyses, citation
  analyses, direct recommendations of other users, or a surprise
  mechanism. Users could choose among the above-mentioned
  search criteria or combine them in order to find the one fitting
  their needs best.
  
  Moreover, comfortable tools for information navigation and the visual browsing of query results would   be available. Finally, users may choose to be alerted of new information of the kind they
  like. Further statistical evaluations could provide a trend
  analysis, identify new users or subjects, emerging fields, and collaboration
  clusters. 

\item A \textbf{networked knowledge manager} would be in charge of
  linking together and updating all pieces of relevant information that were
  already discovered. References and citations of a scientific paper should be
  immediately accessible. Figures and underlying empirical data could be made 
  directly downloadable, while memorizing (and displaying) the original source.

\item A \textbf{shared context-aware reputation system} would
  implement mechanisms to store, exchange, modify, convert and
  share reputation points within decentralized communities.
  Reputation would result from the evaluation of citations and a
  rating system (see Secs. \ref{Rating}, \ref{Reputation}, \ref{Classical}, and \ref{New}). 
  Compared to current systems, raters would be rated as
  well, and their weight would depend on their own
  reputation. Moreover, reputation would be measured on multiple
  scales, allowing one to distinguish, for example, people who have
  many good ideas on different subjects from disciplinary specialists who can elaborate a
  difficult theory, experimental scientists, or people who can write
  good reviews or criticize others' work in a fruitful way. Fairness
  would be rated as well, and content would be classified into
  information (that can be substantiated), opinions, and further
  categories. The novelty of the information would be easily visible,
  as well as the frequency and homogeneity of user responses in
  the different categories. There would be non-anonymous and anonymous
  kinds of ratings and information, but they would be separately
  evaluated and clearly marked. Attempts to manipulate the ratings would
  be sanctioned by reducing the reputation values.

  When rating on multiple scales and recording the identity codes of
  raters (in case of non-anonymous votes), reputation could be evaluated
  in an individualized way. Therefore, the reputation of some
  information or of the person who generated it could vary from one user community
  to another, depending on their customized settings. In this way, 
  communities could develop their own quality standards.
    
\item An \textbf{integrated micro-credit system} to provide incentive
  schemes that reward scientists for their personal contributions. 
  Credits could be earned by certain activities (e.g. rating, reviewing,
  commenting, etc.), and they would be lost in case of vandalism, 
  lack of participation, or spamming etc.
   
  The virtual credits would be primarily spent within the IA system. For
  example, recipients of e-mails could set a price for
  spending their time on them, which may depend on how busy they are, 
  or on the categories of the persons or companies contacting them. 
  Such micropayments should make sure that time budgets of scientists
  will be better used.

  The micro-credit system is also important to reward people for participating
  in crowd sourcing activities (mentioned before). It would therefore be good
  to foresee mechanisms which would allow one to
  convert virtual money raised by this microcredit system into travel grants or
  project funding, for example. Micro-credits could also be used to obtain
  premier information services from the information discovery and retrieval system
  (see Sec. \ref{RecSys}). 
  
  \item A \textbf{virtual education module} should support
    interactive scientific presentations from home without setup or
    travel times. Presentations would be recorded and could be played
    at any convenient time, allowing one to download related
    materials, make notes or comments, or ask questions. Notes would be
    easily searchable and related to each other via tags which
    would be extracted by the system automatically, while users could
    add further tags. Furthermore, notes could be easily shared with
    selected friends or colleagues. Besides lectures, scientists would
    use podcasts to explain their research. 
     
    Eventually, lectures could more and more become like serious, interactive 
    computer games, in which students would interactively explore virtual physical,
    biological, chemical or sociological worlds. The next level in these games would
    be reached, if enough understanding and reputation points have been collected.
    Such educational games could stimulate the imagination, ambition, and learning
    of students and the interested public.
 
\item Finally, a \textbf{privacy settings panel} should offer an intuitive
  interface to regulate and specify at a fine-grained level of detail,
  what data to share with whom, how, for how long, and under what conditions 
  (see \cite{VisioneerCrisis} for more details).

\end{itemize}

\section{New strategies to support a sustainable technological development}

\subsection{Rating Systems and eGovernance}

The general principle of individualized services using a community-based quality evaluation, where not only objects are rated, but raters are rated as well (defining their ``reputation'' and influence/weight), can be transferred to other application areas.
Examples would be new evaluation procedures for project proposals, policies, or public and private services (including the administrative, transport, and health sectors). These issues will
be discussed in more detail in the following.

In particular, rating and reputation systems could be used to support a sustainable
technological development. If the dimensions of the rating scales are properly chosen,
an eGovernance system results. eGovernance can promote positive externalities in several areas, from politics, to business and science. The implementation of eGovernance solutions requires interfaces to set up opinion polls on a variety of subjects in an easy way. Standpoints of different stakeholders would be marked accordingly, and attempts to purposely cheat the system would be sanctioned. Fine-grained, semi-instant and automated polls could allow policy-makers to identify people's current factual and normative projections of the world, i.e. what they are expecting for the future and how they would like it to be. Obviously, participation in such polling activities would be voluntary. It could be coupled to the right to vote and viewed as materialization of it. 

eGovernance opens up many new opportunities. Politicians could get quick feedback on a certain issue from a large number of people in order to 
\begin{itemize}
\item get to know their preferences among certain alternative solutions, 
\item get feedback during the implementation of decisions taken, and
\item evaluate long-term effects and the degree of satisfaction, even years after a policy measure has been implemented.
\end{itemize}

\subsection{Towards indices of human well-being}

The rating and reputation approach could, in particular, help to replace the gross national product (GDP) by better indices oriented at human well-being and sustainability, which have been hard to implement in the past due to difficulties in measuring these. Classically, increasing the gross national product (GDP) has had a great importance in political agendas in the past decades. However, it becomes more and more visible that increasing economic output alone is not sustainable. Therefore, scientists think about better indices to measure human development already for some time. The most prominent document in this connection is probably the manifesto on ``Measuring Economic Performance and Social Progress'' \cite{CommisionSocialProgress}. 
The author list of this report includes several Nobel prize winners such as Joseph Stiglitz, Armatya Sen, Kenneth Arrow, and Daniel Kahnemann. Here are some excerpts from a summary of it \cite{WorldChangingStiglitzManifesto}.

``The seemingly bright growth performance of the world economy between 2004 and 2007 may have been achieved at the expense of future growth. It is also clear that some of the performance was a `mirage', profits that were based on prices that had been inflated by a bubble.

The whole Commission is convinced that the crisis is teaching us a very important lesson: those attempting to guide the economy and our societies are like pilots trying to steering a course without a reliable compass. The decisions they (and we as individual citizens) make depend on what we measure, how good our measurements are and how well our measures are understood. We are almost blind when the metrics on which action is based are ill-designed or when they are not well understood. For many purposes, we need better metrics. ... the time is ripe for our measurement system to shift emphasis from measuring economic production to measuring people's well-being. ...
To define what well-being means, a multidimensional definition has to be used. Based on academic research and a number of concrete initiatives developed around the world, the Commission has identified the following key dimensions that should be taken into account. At least in principle, these dimensions should be considered simultaneously:

\begin{itemize}
\item[i.] Material living standards (income, consumption and wealth); 
\item[ii.] Health; 
\item[iii.] Education; 
\item[iv.] Personal activities including work;
\item[v.] Political voice and governance;
\item[vi.] Social connections and relationships; 
\item[vii.] Environment (present and future conditions); 
\item[viii.] Insecurity, of an economic as well as a physical nature.
\end{itemize}

All these dimensions shape people's well-being, and yet many of them are missed by conventional income measures. Steps should be taken to improve measures of people's health, education, personal activities and environmental conditions. In particular, substantial effort should be devoted to developing and implementing robust, reliable measures of social connections, political voice, and insecurity that can be shown to predict life satisfaction. 
Statistical offices should incorporate questions to capture people's life evaluations, hedonic experiences and priorities in their own surveys.'' For this, the above proposed rating system could be very useful.

\subsection{The economics of happiness}

It is remarkable that many of the above mentioned dimensions correlate very well with ``happiness''. While economic research has paid attention to it only recently \cite{RePEc:mtp:titles:0262062771}, happiness indeed plays a prominent role in the American Declaration of Independence on July 4, 1776. The second paragraph starts with the statement: ``We hold these truths to be self-evident, that all men are created equal, that they are endowed by their Creator with certain unalienable Rights, that among these are Life, Liberty and the pursuit of Happiness.''
\par
Happiness as a concept and goal of life is being rediscovered, recently. For example, a study of Deutsche Bank Research \cite{DeutscheBankHappiness} reached the following conclusions:
\begin{itemize}
\item Germany as one of best ranked countries in the GDP statistics has a less happy population as compared to many other countries ranked far behind Germany.
\item Happy societies tend to be characterized by a high level of trust, a low level of corruption, a low level of unemployment, a late retirement, a small shadow economy, a high education, many freedoms, a high income, and more children.
\end{itemize}

In other words, happy societies seem to be more sustainable in the long run. Based on suitable rating scales, the rating system of the Innovation Accelerator could be used to measure the relevant dimensions of happiness, which certainly entail, but at the same time go beyond, monetary reward \cite{ElizabethW.Dunn03212008,PriceHappiness}. The outcomes of such measurements could be used to orient policy-makers, i.e. give them a better compass than GDP was.\footnote{It is worth noting that, recently, a ``Bank of Happiness'' \cite{BankHappiness} has been established, where people can trade good deeds instead of companies' stocks. Other initiatives point into the same direction \cite{doogood}.} 

\subsection{New incentive systems}

One important component in the creation of a more sustainable economics could be the creation of new incentive systems which go beyond sheer profit maximization. In fact, it is well-known that individuals respond to non-monetary incentives as well \cite{RePEc:mtp:titles:0262062771,citeulike:969736,ClaEtAl98}, such as prizes or medals or just compliments. The current science system is, in fact, a very good illustration of this principle. Authors spend a lot of time---and spare time---on creating manuscripts which they usually publish without taking money for this. In the same way, they participate in review, dissemination and administration activities without monetary compensation. One of the main motivations for this seems to be the reputation that they can gain among their peers (their colleagues). The existence of ranking scales (such as citation scales) is often enough to stimulate their ambition. Therefore, other ranking scales, something like area-specific ``hit parades'', could probably create suitable incentives to engage in voluntary, socially beneficial activities in various areas of society and economy. Multi-dimensional reputation scales (see Sec. \ref{Reputation}) could serve this purpose. For example, it is well known from computer games (and multi-player online games) that individual points or rankings on competitive scales can be very efficient incentives for people, stimulating their ambition and making them invest a lot of time. 

\section{New ways to reach societal benefits and
  respond to industrial needs using ICT}

Quite obviously, individualized services using a community-based quality evaluation, where not only objects are rated, but raters are rated as well, can be also used to assess product quality, to rate architectural designs, or to support the decision-making in companies. This only requires a suitable choice of the rating scales (the price, the quality of ingredients, durability, production conditions, environmental impact, etc.). Rating and information services could also be based on mobile phones and WLAN architectures, e.g. when consumers buy products in a shop and want to have product information or want to rate products.
\par
Such technologies would allow companies to produce goods which come closer to their clients' dreams. This apparently requires to  systematically analyze and consider meaningful customer opinions (e.g. wish lists) in the design of all stages of the product life cycle. However, not only could companies scale and speed up their early R\&D activities dramatically by using crowd sourcing and other tools offered by the Innovation Accelerator. The feedback loops created by its rating systems would make it possible as well to realize the vision of participating consumers, or even consumers contributing to the production process (so-called ``prosumers'' \cite{citeulike:1261291}). 

It is certainly not our task here to define the policies and technologies which serve the purpose of supporting economic progress best, while promoting social well-being and a sustainable environment. We believe that there is no simple and unique answer to this question. Nonetheless, there is a responsibility to prepare adequate tools to put future policy-makers into the position to take decisions based on the best available knowledge. ``The world is filled with minds that can
contribute. And when the information is shared, we can just move faster'' \cite{ScienceCommonsPromo}.

\section{Summary and final considerations}

The ``Lisbon agenda'' aims at creating a knowledge-based economy in Europe driven by innovation. In order to achieve this,
it is necessary to re-invent innovation, in particular the way how science is performed at academic institutions. It is apparent that
\begin{itemize}
\item the evaluation system does not work well anymore---it has passed its optimum and becomes more and more a burden,
\item a new funding system is needed, which could be based on rewarding previous performance,
\item for this, one needs to develop multi-dimensional and fairer performance measures, as current indices
are too biased and unfair,
\item new publication, communication, coordination, and co-creation concepts are needed to optimize the innovation rate 
and the dissemination of the best ideas. 
\end{itemize}
While various activities in this direction have started, there is still a long way to go. However, the benefits of 
novel ICT-based systems are expected to be  large not only for science, but also for societies and economies.
For example, multi-factorial reputation and recommender systems could connect producers and consumer
communities more closely with each other, simplifying everything from
crowd sourcing and prediction markets over customized services and
personalized education to participatory production and consumption. 
\par
The need for an Innovation Accelerator becomes particularly obvious when analyzing some institutional obstacles 
in the socio-economic and other research fields that the financial crisis has revealed. 
Despite its inherent logic, the current economic crisis and its course of
events has not been well anticipated (cheap credits, US real estate bubble, quant meltdown,
default of Lehmann Brothers, liquidity crisis, bank bankruptcy
cascade, defaults of companies, mass unemployment, public spending
deficits, instability of European currency, public saving plans, ...). 
This suggests that current mainstream theories do not describe such phenomena
well enough. Considering that economic crises implies a loss of property and security for many people and potentially serious impacts on their lives, it seems required to think about beneficial institutional changes. 

\subsection{The education system}

Some of the institutional problems that need to be overcome concern the education system. In many scientific fields, including various socio-economic sciences, academic curricula seem to lag considerably behind the scientific state-of-the-art. For example, in economics, rather than promoting principles like sustainability and fairness, economics still confronts students predominantly with a world view of profit
maximization, while scientists have revealed since a long time
that humans responds to other, non-monetary incentives as well \cite{RePEc:mtp:titles:0262062771,citeulike:969736,ClaEtAl98}, and that monetary incentives can damage
voluntary commitments \cite{GneezyRustichini}. This biased educational approach results
from the leading paradigm of the ``homo economicus'', i.e. of the ``perfect
egoist'', which is promoted by most economics text books and many research articles 
despite of contradicting evidence (see Ref. \cite{BaliettiEconophysics}
for a more detailed discussion of this point). It is quite interesting to ask why 
the discoveries of Nobel prize winners
in economics, like Kahnemann and Tversky, Selten,
Schelling, Akerlof, Stiglitz, Krugmann, or Ostrom, to give just a few
examples, have not managed to change this paradigm over the decades. Most likely, this is
a consequence of the difficulty to obtain suitable data to test socio-economic 
theories in the past. However, university courses also need  to change in other respects.
They should be significantly adapted, considering the revolution in the
area of information technology that the world has seen and the
significant progress in areas such as complex systems modeling and
computer simulation, or data mining. 

\subsection{The recruitment system and incentive structures}

One of the central question is why we do not see a larger degree of
innovation and change as compared to other fields. In
physics, for example, classical mechanics was
replaced by quantum mechanics and relativistic mechanics. Moreover,
it has been complemented by electrodynamics, statistical mechanics,
and a varity of other fields. New, interdisciplinary fields like
bio-, traffic-, econo- or socio-physics have been created. In the
past few decades, there have been a number of new research focuses
such as superconductivity, nanoscience, spin glasses, neural
networks, chaos theory, or network science. It appears that not all 
scientific fields have the same innovation dynamics.
\par
One of the underlying problems may be the prevalent recruiting
system. In order to become professor of economics, one must have
published in the leading peer-reviewed journals of economics. The
number of these A-rated journals is relatively small, and the number
of articles is limited. The selection of papers that are finally
published in these journals is carried out by a relatively small
number of people who are following more or less the same economic
paradigm(s). This creates similar problems as known from oligopoles.
Not only can an ``old boys club'' control the market of publications
by the way, in which manuscripts are selected. Ambitious junior
scientists also have to subject their research subject and research
approach to the preferences of these journals, otherwise they are
punished with the ``academic death penalty'' of being chance-less in
the competition for chairs in economics.

\subsection{The publication system}

Many scientists are complaining about their journals:

\begin{enumerate}

\item for the long publication delays (in some fields, the publication of scientific
manuscripts in case of acceptance typically takes 2 or 3 years as compared to 4 to 6 months in
physics or biology, and it takes even longer in case of rejection),

\item for the narrow selection of manuscripts and the little freedom to come up with innovative ideas (quality control also introduces a certain 
degree of ``censorship'' into the scientific system, in contrast to the way the public media work).

\end{enumerate}

The level of pluralism seems to be largely dependent on the field.
In microeconomics, for example, there is just one predominant
approach, which is the paradigm of the ``homo economicus''. In
macroeconomics, there are just two, namely Keynesianism and
neoclassical economics, and both of these are not fully supported empirically.
Nevertheless, they are the prevailing approaches informing today's policies. 
\par
For systems as complex as economies, a pluralistic
modeling approach would certainly be more appropriate \cite{PluralisticModeling}.
Today, it appears that innovation is often slowed down by requiring that new
results should be consistent with previous ones. Kuhn's work on scientific revolutions \cite{KuhnScientificRevolutions}
suggests that the evolution of knowledge is not gradual, but
characterized by paradigm shifts. It appears, however, that such paradigm
shifts have not happened in certain fields for a long time.
\par
A further obstacle to innovation seems to be the focus on purely
analytical results. This restricts research mainly to relatively
simple models. It is therefore hard to see what can happen in
complex systems with strong non-linear interactions, random
influences, spatio-temporal and network dynamics, and heterogeneity.
We think that many of the remaining scientific challenges these days
cannot be solved without the use of computers. The tools and
instruments dominating in certain scientific fields do not appear to be 
fully sufficient to understand the complexity of the studied systems \cite{BaliettiEconophysics}.
While areas like physics and biology are using the most powerful computers to simulate their systems and
turn data into knowledge by sophisticated data mining concepts, these
methodologies have not spread into all scientific fields so far.
It is almost as if we would not use all our senses to get a picture
of the world.

\subsection{The research approach}

While the scientific problems to be solved are big, average research
teams are very small. This promotes the specialization on details, while
systemic studies, e.g. the study of systemic risks, are rare. In fact,
certain fields seem to be pretty much fragmented. This situation seems largely due to a lack of
\begin{enumerate}
\item methods from complexity science,
\item computational modeling,
\item empirical and experimental data,
\item engineering-like solutions and tested alternatives.
\end{enumerate}

In areas dominated by mathematical analysis, multi-disciplinary
collaborations seem to be rare, although
behavioral studies, statistical physics and complex systems theory,
agent-based modeling, machine learning, artificial intelligence,
systems design and advanced testing would be highly relevant to
advance the knowledge and design in a number of fields involving human behavior. 
\par
A problem that comes with a largely axiomatic approach (which can be
quite successful, of course) is the lack of orientation at the real world (data and problems). 
Experimental and data mining approaches are coming up in some fields
only recently, and suitable experimental and measurement procedures still
need to be developed. Although difficult, it does not seem to be out
of reach. Moreover, many analytical results are only achievable with approximations. 
Consequently, the quality of an approximation must be either
tested against a quasi-exact numerical solution or against
experimental data sooner or later to identify the limitations of a model.
Again, the collaboration with other disciplines should be fruitful here.

\subsection{The knowledge creation cycle}

One must be aware that the remaining puzzles concerning the behavior of complex systems
can probably not be solved with the methods used in the past, but require the use of new 
approaches. When resulting from non-linear feedback effects, which are
typical for networked systems with strong interactions, they often have 
counter-intuitive explanations. Most large real-life systems
are of exactly of this kind. Such non-linear interactions often give 
rise to self-organization and emergent
phenomena, i.e. the sudden appearance of new properties (innovations
are typical examples). As the system behavior is difficult to predict, 
control may be an illusionary concept (see Refs. \cite{citeulike:3975610,IRGC} for a detailed discussion). It is essential here to
underline that self-organization and emergence cannot be understood
with equilibrium or linear models. They require non-linear 
dynamical models, while many empirical studies 
use (multi-variate) linear models, i.e. statistical approaches that 
cannot reveal the non-linear laws underlying complex systems.
\par
Scientific progress requires a number of different steps, including non-linear data analysis, mathematical modeling, computer
simulation, and finally, optimization, management, control or systems
design. Specialization often prevents researchers to be engaged in
all steps of this scientific knowledge creation process. Therefore,
working in larger, multi-disciplinary teams appears to be a necessary
development, particularly when considering the long publication times
at present.
\par
It must be underlined that scientific knowledge creation is neither a
one-step process nor a linear progress, but is better imagined as a
networked system with feedback loops. It seems natural, for example,
to start of with simple models to explain certain stylized facts. As
good data sets become available, e.g. through lab experiments,
empirical analyses, or technological developments, models can be
calibrated, tested, verified, falsified, or improved. Usually, the
data analysis reveals new facts that require a consequent modification or
extension of the model(s). Therefore, a plurality of alternative
models can speed up the development of realistic models \cite{PluralisticModeling}. Some day,
models become so good that they can reproduce the majority of
observations. It becomes possible then to apply them to practical problems. 
\par
Realistic challenges on the other hand call for specific
types of models. For some questions, aggregate (``macroscopic'')
models may be more adequate, for other questions,
``microscopic'' (e.g. agent-based approaches) may be more suitable.
Moreover, overseeing the collection of stylized facts that the
interaction between data analyses and computational predictions has
revealed, it becomes possible to simplify realistic models in a way
that allows an analytical understanding at the cost of quantitative
accuracy. All these different approaches and steps are needed to make
scientific progress. If the interaction between these steps is
obstructed, scientific progress may be slow or impossible. For
example, when requiring realistic models right away, or when
narrowing down the number of models too much in an early stage of the
development of a field, scientific progress will be hardly possible.
Unfortunately, it seems that the interaction between different
disciplines involved into the research process currently does not
work very well in some scientific fields, i.e. there are missing links or at least
weak ones, as compared to other scientific fields. This is not only a
problem of research traditions, but also an institutional problem,
which needs to be addressed and can be overcome.

\subsection{Suggestions for institutional changes}

In the following, we suggest some measures that can be taken to
overcome the problems mentioned before:
\begin{enumerate}
\item The content of university studies should be adapted to the general
scientific and technological progress (state-of-the-art). Students should have courses
in programming languages and the use of data mining, computer
simulation, and visualization tools. Moreover, the study contents should be
more interdisciplinary. It would certainly advantageous to get a basic overview 
of complexity science, the social and natural sciences (e.g. network science) and 
engineering (e.g. cybernetics).
\item The preselection of faculty members should better be based on relative
citations rather than the (impact factor of the) journals one has
published in. However, even more important than this is the
assessment of the content and quality of publications written,
particularly their intellectual depth and level of innovation. Therefore,
reputation values as determined by suitable rating systems would be 
important complementary information.
\item Recruitment committees should reward interdisciplinary research
projects as well as data-oriented and problem-oriented research
rather than the use of certain methods.
\item The publication system should become more efficient (in terms of
publication times), more transparent for innovative (heterodox)
approaches, and pluralistic (scientific convergence should happen
based on successful testing, not through selective control).
\item Replacing journal editors by open-minded, multi-disciplinarily
oriented editors may help. Accepting a certain percentage 
of heterodox papers (say 20--30\%) would stimulate innovation and support
plurality rather than narrow, unidirectional research. Journals
should also support formats such as comments and replies, e.g.
through a discussion section or by providing a corresponding
functionality at the internet portals of the electronic journal
version. Furthermore, methodological contributions and contributions
from related fields, such as econophysics, for example, should be
sporadically accepted. Such contribution could be marked as guest
contributions, comments, opinions, etc., depending on the respective
kinds of contributions. Moreover, new, multi-disciplinarily oriented journals
should be launched, using the various possibilities offered by Web2.0 platforms
(see Sec. \ref{Science2}). 
\item Funding agencies should reward collaborative, interdisciplinary,
goal-driven research. Furthermore, incentives for computational,
experimental and data-mining studies should be provided.
\item Research agencies should create special funding schemes and
evaluation procedures for non-standard, high-risk research. For
example, scholarships that allow researchers to travel and to visit
different institutions and departments would be useful. Moreover,
large research grants would encourage universities to recruit faculty
members who were awarded with them, particularly when they come with
overheads for the hosting institution under the condition that a
permanent position is offered.
\item The research cycle should be improved, creating missing links and feedback loops.
The development of an Innovation Accelerator, as outlined in Sec. \ref{QJournal}, would provide the
right communication, coordination, and co-creation tools for this.
\end{enumerate}

\subsection{Final considerations: Publishers as future information brokers?}

It is quite obvious that both, the scientific system and the publication system will face major paradigm shifts. In fact, we are expecting ``more change in the next 50 years of science than in the last 4 hundreds years of inquiry'' \cite{KellyFutureScience}. For example, the journal business of publishers may change considerably. Journals could largely be replaced by self-organizing information systems, using reputation and crowd sourcing methods. The future role of publishers may be that of knowledge scouts or knowledge brokers, earning mainly on discovering and connecting information rather than on disseminating information. Thereby, they could play an important role for future knowledge transfer, connecting science with business and politics much better than it has happened in the past. In order to make this knowledge transfer successful, however, it will be crucial to find fair ways of sharing profits proportionally with the originators of ideas and inventions. 

\subsection*{Acknowledgements}

The authors are grateful for financial support by the Future and Emerging Technologies programme FP7-COSI-ICT of the European Commission through the project Visioneer (grant no.: 248438).
They would also like to thank for stimulating discussions, feedback on the
manuscript and contributions to the Visioneer wiki: Santo Fortunato, Andras L\"orincz, Jorge Lou\c{c}\~a, Sergi Lozano, Matus Medo, Mario Paolucci, Andrea Scharnhorst, Magda Schiegl and Gabriele Tedeschi. 

\appendix

\section{Pluralistic indices promoting individuals talents}\label{pleval}

One important question is, how the performance values $X_i$ on multiple scales $i$ can be translated into one scale. 
Traditionally, this is done by weighting each criterion $i$ with a certain factor $w_i$. This results in average values
\begin{equation}
x = \sum_i w_i x_i \, , 
\end{equation}
where $x_i = X_i / \langle X_i \rangle$ is the value $X_i$, which has been scaled by the average performance $\langle X_i \rangle$. The individual values $x$ can ordered on a one-dimensional scale, i.e. ranked. Such an approach, however, promotes average performance rather than excellence, as the latter is typically characterized by extreme values on one or a few rating scales, but not on all of them. In order to reward individual rather than average talent, two methods seem to be interesting:
\begin{enumerate}
\item Similar to the world of sports, one could classify different leagues (A-league, B-league, C-league, etc.).
The A-league could consist of those people, who are among the $y$\% best on one scale, or among
the 2$y$\% best on two scales, or among the 3$y$\% best on three scales among all competitors (where one could, for example, choose $y=10$). The B-leage would consist of the candidates, who would not be good enough to meet the standards of the A-league, but would meet them for a suitably chosen, higher value of $y$. One could proceed similarly with the C-league.
\item Political decision-makers could choose the weight they attribute to each criterion, say $w_1 = 0.35$, $w_2=0.25$,
$w_3 = 0.25$, and $w_4=0.15$, where criterion $i=1$ could, for example, be scientific excellence, $i=2$ industrial relevance, $i=3$, societal or political relevance, and $i=4$ dissemination. An index, which would be favorable with respect to individual talent, would be
\begin{equation}
 y = \sum_i w_i x_i + 0.1 (y_1 + y_2 - y_3 - y_4) \, , 
\end{equation}
where the values $y_i$ correspond to the values $x_i$, sorted according to their size in descending order. This formula overweights particular individual talents, i.e. it gives everyone a fair chance and rewards excellence in specific areas.
\end{enumerate}
Both indices overcome some a number of problems of the ranking methods that are primarily used today (one-dimensional ones or those averaging in an individually non-differentiated way). Nevertheless, it is advised to use them only for preselection and give human experts the final word in the performance assessment, as automated procedures are not perfect and tend to overlook relevant factors (such as family- or health-related ones).

\section{Noteworthy 2.0 Collaborative Initiatives}\label{web2.0}

\subsection{Crowdsourcing}

\begin{itemize}

\item \textbf{Innocentive} 

Innocentive offers a web dashboard where companies can post scientific challenges seeking for alternative solutions. Successful problem solvers get rewarded with monetary prizes.  

\url{http://www.innocentive.com/}

\item \textbf{Stack Overflow}

Stack Overflow is a collaboratively edited question and answer site
for programmers.

\url{http://stackoverflow.com/}

\item \textbf{Stack Exchange}

Stack Exchange is a network of free, community-driven Q\&A sites. We
highlight and aggregate the best recent contents from our entire
network here. Area 51 is a side-component of the website which allows
the creation of new Q\&A sites through an open and democratic process.

- \url{http://stackexchange.com/}\\
- \url{http://area51.stackexchange.com/}

\item \textbf{Toolbox}

Toolbox provides a web solution for knowledge sharing within
topic-centric communities of professionals.

\url{http://www.toolbox.com/}

\item \textbf{Corporate Executive Board}

The Corporate Executive Board delivers authoritative data and tools, best practice research, and
peer insight to the leaders of the world’s great enterprises.

\url{http://www.executiveboard.com/}

\item \textbf{43 Things}

On 43 Things people can post a list of 43 goals to reach in their lives, from watching a space shuttle launch to grow my own vegetables. It entails a social space where people answer questions regarding how they achieved specific goals.

\url{http://www.43things.com/}

\item \textbf{KickStarter}

Kickstarter offers an intuitive web interface for performing fund-raising for artistic, scientific and engineering projects.

\url{http://www.kickstarter.com/}

\item \textbf{Taking It Global}

TakingITGlobal tries to pull together people from all the world towards globally relevant issues.

\url{http://www.tigweb.org/}

\item \textbf{Bank of Happiness}

The purpose of the Bank of Happiness is to promote non-monetary values in order to help people find their way back to the deeper values. It offers a Web 2.0 site from which people can exchange good deeds for free.

\url{http://www.onnepank.ee/}

\end{itemize}

\subsection{Content Aggregation}

\begin{itemize}

\item \textbf{Research Blogging}

ResearchBlogging.org is a system for identifying thoughtful blog posts about peer-reviewed research. Users create properly-formatted research citation for the journal articles with an automated citation generator, then they paste a special code into their blog entry and an automated aggregator finds their post and publishes it on the front page.

\url{http://researchblogging.org/}

\item \textbf{Digg}

Digg allows to discover and share content on the Internet, by submitting links and stories, and voting and commenting on submitted links and stories.

\url{http://digg.com/}

\item \textbf{Redditt}

Redditt allows to discover and share content on the Internet, by submitting links and stories, and voting and commenting on submitted links and stories.

\url{http://www.reddit.com/}

\item \textbf{Delicious}

A social bookmarking tool to share and tag internet references.

\url{http://www.delicious.com/}

\item \textbf{Yahoo Pipes}

Pipes is a powerful composition tool to aggregate, manipulate, and mashup content from around the web.

\url{http://pipes.yahoo.com/pipes/}

\item \textbf{StumbleUpon}

Stumble Upon allows the discovery of new web content on the basis of powerful recommendation system applied to a vast user-base.

\url{http://www.stumbleupon.com/} 

\end{itemize}

\subsection{Project Management}

\begin{itemize}

\item \textbf{Stakesource}

Social networking tool that automatically identifies and prioritises
the stakeholders for your projects, engages with the stakeholders, and
understands their needs. Crowdsourcing, collaborative filtering and
prioritised lists are among the tools available here.

\url{http://www.stakesource.co.uk/}

\item \textbf{Scrum}

Scrum is an agile framework for completing complex projects. Scrum
was originally formalized for software development, but it suits
well any complex and innovative project.

\url{http://www.scrumalliance.org/}

\item \textbf{SharePoint Workspace 2010}

Formerly known as Groove, it is a collaborative work platform used by
Microsoft.

\url{http://office.microsoft.com/en-us/sharepoint-workspace/}

\end{itemize}

\subsection{Measuring Scientific Progress}

\begin{itemize}

\item \textbf{Scholarometer}

Scholarometer(beta) is a social tool to facilitate citation analysis
and help evaluate the impact of an author's publications.

\url{http://scholarometer.indiana.edu/}

\item \textbf{Liquid Pub}

This project promotes the notion of Liquid Publications which are evolutionary, collaborative, and composable scientific contributions. Based on lessons learnt from open source software development and from Web 2.0 applications, it promotes a software platform for collaborative evaluation of knowledge artifacts.

\url{http://liquidpub.org/}

\item \textbf{HistCite}

HistCite helps science professionals to make better use of the
results of their searches of the Web of Science. HistCite lets you
analyze and organize the results of a search to obtain various views
of the topic’s structure, history, and relationships. 

\url{http://www.histcite.com/}

\item \textbf{Phys Author Rank Algorithm}

Phys Author Rank Algorithm is a website where physicists can check the
evolution of their own scientific rank. Scientific rank is calculated
using the Science Author Rank Algorithm on a weighted author citation
network.

\url{http://www.physauthorsrank.org}

\item \textbf{Citebase}

Citebase Search is a semi-autonomous citation index for free,
online research literature. Citebase contains articles from physics,
maths, information science, and (published only) biomedical
papers. Currently, only an experimental demo is working.

\url{http://www.citebase.org/}

\item \textbf{Publish or perish}

Publish or Perish is a software program that retrieves and analyzes
academic citations from Google Scholar. It obtains the raw citations,
analyzes them, and presents several statistics.

\url{http://www.harzing.com/pop.htm}

\item \textbf{Scholar Index}

Scholar index is a free service to query Google Scholar and to compute
and visualize the corresponding h-index and other metrics.

\url{http://interaction.lille.inria.fr/~roussel/projects/scholarindex/}

\item \textbf{SPIRES}

SPIRES is an eprint repository for particle and nuclear physics contributions. It displays a number of additional information about citations and indexes.

\url{http://www.slac.stanford.edu/spires/}

\item \textbf{SciVal}

Based on co-citation analysis, SciVal by Elsevier  offers of an overview of institutions' research performance, indicating growing areas of multidisciplinary strength and identifying key competitors and potential collaborators. Not free of charge.

\end{itemize}

\subsection{Scientific References Management}

\begin{itemize}

\item \textbf{Citeulike}

Citeulike is a free service for managing and discovering scholarly
references. It allows one to easily store references, to find new ones
through recommendations and to share references with peers.

\url{http://www.citeulike.org/}

\item \textbf{Zotero}

Zotero is a free, easy-to-use Firefox extension for collecting, managing, citing and sharing your research sources. 

\url{http://www.zotero.org/}

\item \textbf{Mendeley}

Mendeley is a free research management tool to
organize, share and discover research papers.
Improved services are available for charge.

\url{http://www.mendeley.com}

\item \textbf{Connotea}

Free online reference management for all researchers, clinicians and
scientists. It saves and organizes links to your references and shares
them with colleagues.

\url{http://www.connotea.org/}

\item \textbf{Bibsonomy}

BibSonomy is a tags-based system for sharing and organizing lists of scientific references.

\url{http://www.bibsonomy.org/}

\item \textbf{CiteSeerX}

CiteSeerX is a scientific literature digital library and search engine that focuses primarily on the literature in computer and information science. Rather than creating just another digital library, CiteSeerX attempts to provide resources such as algorithms, data, metadata, techniques, and software that can be used to promote other digital libraries. The code of the engine is open source and downloadable.

\url{http://citeseerx.ist.psu.edu/}

\item \textbf{World Wide Science}

WorldWideScience.org is a global science gateway comprised of national and international scientific databases and portals. WorldWideScience.org provides one-stop searching of databases from around the world. A beta feature adds real-time translation of multilingual scientific literature.

\url{http://worldwidescience.org/}

\item \textbf{Faculty of 1000}

Faculty of 1000 Biology and Medicine are authoritative online services in which over 5,000 leading researchers and clinicians share their expert opinions by highlighting and evaluating the most important articles in biology and medicine.

\url{http://f1000.com/}

\item \textbf{SAO/NASA ADS}

Managed by NASA, it supplies detailed bibliographic information about physics papers. Users can obtain on a record by record basis all the information available about a particular bibliographic entry (including the bibliographic code, title, authors, author affiliations, journal reference, publication date, category, comments, origin, keywords, and abstract text when available). 

\url{http://adsabs.harvard.edu/physics_service.html}

\end{itemize}

\subsection{Spatial Visualization of Science}

\begin{itemize}

\item \textbf{Living Science}

Living Science is a real-time global science observatory based on
publications submitted to arXiv.org. It covers daily
submissions of publications in areas as diverse as Physics, Astronomy,
Computer Science, Mathematics and Quantitative Biology. Currently,
contents are dynamically updated every day. Living Science is an analysis
tool to identify the magnitude and impact of scientific work worldwide.

\url{http://www.livingscience.ethz.ch/}

\item \textbf{AuthorMapper}

AuthorMapper searches journal articles and book chapters and plots the
location of the authors on a map.

\url{http://authormapper.com/}

\item \textbf{Eigenfactor}

Eigenfactor offers interactive visualizations based on the Eigenfactor\texttrademark Metrics and hierarchical clustering to explore emerging patterns in citation networks.

-\url{http://eigenfactor.org/}\\
-\url{http://well-formed.eigenfactor.org/}\\

\item \textbf{Visualizing Arts and Humanities Citation Index}

The site displays the position and environment of every individual journal in A\&HCI (2008) based on their similarities in citation patterns.

\url{http://vks2.virtualknowledgestudio.nl/ahci/index.html}

\end{itemize}

\subsection{Scientific Social Networks and Collaborative Tools}

\begin{itemize}

\item \textbf{Academia.edu}

Academia.edu is a system that manages publications and shows institutions and scientists in a graphical tree representation. It allows to follows other colleagues' work and share one's own publications within a network of trusted peers.

\url{http://www.academia.edu}

\item \textbf{Research Gate}

Research Gate is a professional social network for scientists. It's free of charge.

\url{http://www.researchgate.net/}

\item \textbf{BioMedExperts}

BioMedExperts (BME) is a web platform to allow scientists and
researchers across multiple organizations and nations to share data
and to collaborate.

\url{http://www.biomedexperts.com/}

\item \textbf{ResearcherID}

ResearcherID is a global, multi-disciplinary scholarly research
community.  With a unique identifier to each author in
ResearcherID, one can eliminate author misidentification and view an
author's citation metrics instantly. ResearcherID allows one to search the registry 
in order to find
collaborators, review publication lists, and explore how research results are
used around the world.

\url{http://www.researcherid.com/}

\item \textbf{Science Commons}

Science Commons designs strategies and tools for faster, more
efficient web-enabled scientific research.

\url{http://www.sciencecommons.org/}

\item \textbf{NeuroCommons}

It is a ``proof-of-concept'' project from the Science Commons
Initiative within the field of neuroscience. The NeuroCommons is a
beta open source knowledge management system for biomedical research.

\url{http://www.neurocommons.org}

\item \textbf{Nature Network}

Nature Network is an online network for scientists to discuss
scientific news and events.

\url{http://network.nature.com/}

\end{itemize}

\printbibliography

\end{document}